\documentclass[twocolumn,aps,showpacs]{revtex4}
\usepackage {epsfig}
\usepackage{graphicx}
\usepackage{epstopdf}
\usepackage{amssymb}

\begin{document}
\bibliographystyle{apsrev}
\title{Partial synchronisation of stochastic oscillators through hydrodynamic coupling}
\author{Arran Curran\footnote{Electronic mail: arran.curran@glasgow.ac.uk}}
\author{Michael P. Lee}
\author{Miles J. Padgett}
\affiliation{School of Physics and Astronomy, SUPA, University of Glasgow, Glasgow G12 8QQ, United Kingdom}
\author{Jonathan M. Cooper}
\affiliation{School of Engineering, University of Glasgow, Glasgow G12 8LT, United Kingdom}
\author{Roberto Di Leonardo}
\affiliation{CNR-IPCF, UOS Roma, Dipartimento di Fisica Universit\`a ``Sapienza", I-00185 Roma, Italy}
\date{\today}

\begin{abstract}
Holographic optical tweezers are used to construct a static bistable optical potential energy landscape where a Brownian particle experiences restoring forces from two nearby optical traps and undergoes thermally activated transitions between the two energy minima. Hydrodynamic coupling between two such systems results in their partial synchronisation. This is interpreted as an emergence of higher mobility pathways, along which it is easier to overcome barriers to structural rearrangement.
\end{abstract}
\maketitle

Activated dynamics is ubiquitous in physics, chemistry and biology \cite{kramer,hanggi1990reaction}.  Escape from a stable state occurs via transition events by which a system crosses an energy barrier, falling into a neighboring state. On the small length scales of colloidal particles and biological macromolecules, it is the surrounding fluid which provides the required fluctuations to overcome a barrier and, at the same time, the viscosity needed to relax the system in the second minimum by dissipating energy. With the addition of a spatial potential landscape consisting of two or more minima, such an environment allows a colloidal particle to undergo hops between the minima, provided that any energetic maximum is surmountable by thermal fluctuations \cite{golding, Kotar:2010p4026, BENZI:1981p5914}. Starting from a Fokker-Planck formulation of Brownian motion in bistable potentials, Kramers \cite{kramer} was able to obtain an analytical expression for the hopping rates that depends solely on the local features of the underlying potential. McCann \emph{et al.} \cite{golding} used two separate HeNe lasers to construct a bistable optical landscape where a $600$ $nm$ Silica sphere was confined in a solvent fluid. The sphere hopped back and forth between the traps, providing the ideal representation of Kramers' ideas and elegantly confirmed the link between hop rates and the features of the potential landscape. Since this first quantitive test of Kramers' theory, a number of experiments have been published investigating how the hop rates of stochastic oscillations behave under modulation of the potentials \cite{simon_resonance, goldingPRL, LehmannPRL, maierPRL, Visscher:2009}. By asymmetrically modulating the depth of the potential wells and hence the phase between the forcing and the activated transitions, Simon and Libchaber \cite{simon_resonance} observed synchronization of stochastic oscillations with a period equal to the mean Kramers time.

On this micron scale, both fluctuations and friction display long range hydrodynamic correlations that inevitably couple the dynamics of suspended objects. For example, two colloidal particles in independent optical traps display correlations in their Brownian motions \cite{Meiners:1999p2211}. Whether the same hydrodynamic interaction could result in some degree of correlation in the thermally activated dynamics of nearby bistable systems is still an open and important question. Understanding how hydrodynamics can bias activated hops over energy barriers impacts the study of the slow dynamics of structural rearrangements in colloidal glasses and gels.

In this Letter we use two colloidal spheres trapped in neighboring but separate bistable optical landscapes to reveal a significant difference between symmetric and antisymmetric simultaneous hops. Whilst we measure a significant biasing, the total number of simultaneous hops transpires only to be that expected from two random oscillators. This is due to the stochastic nature of our oscillators, whose dynamics are driven by thermal fluctuations alone. We show that the difference between the number of symmetric and antisymmetric hops reduces as the inverse separation between the two systems, showing that the strength of the phenomenon scales with first order in the hydrodynamic coupling.

\begin{figure}[htb]
\includegraphics[width=8cm]{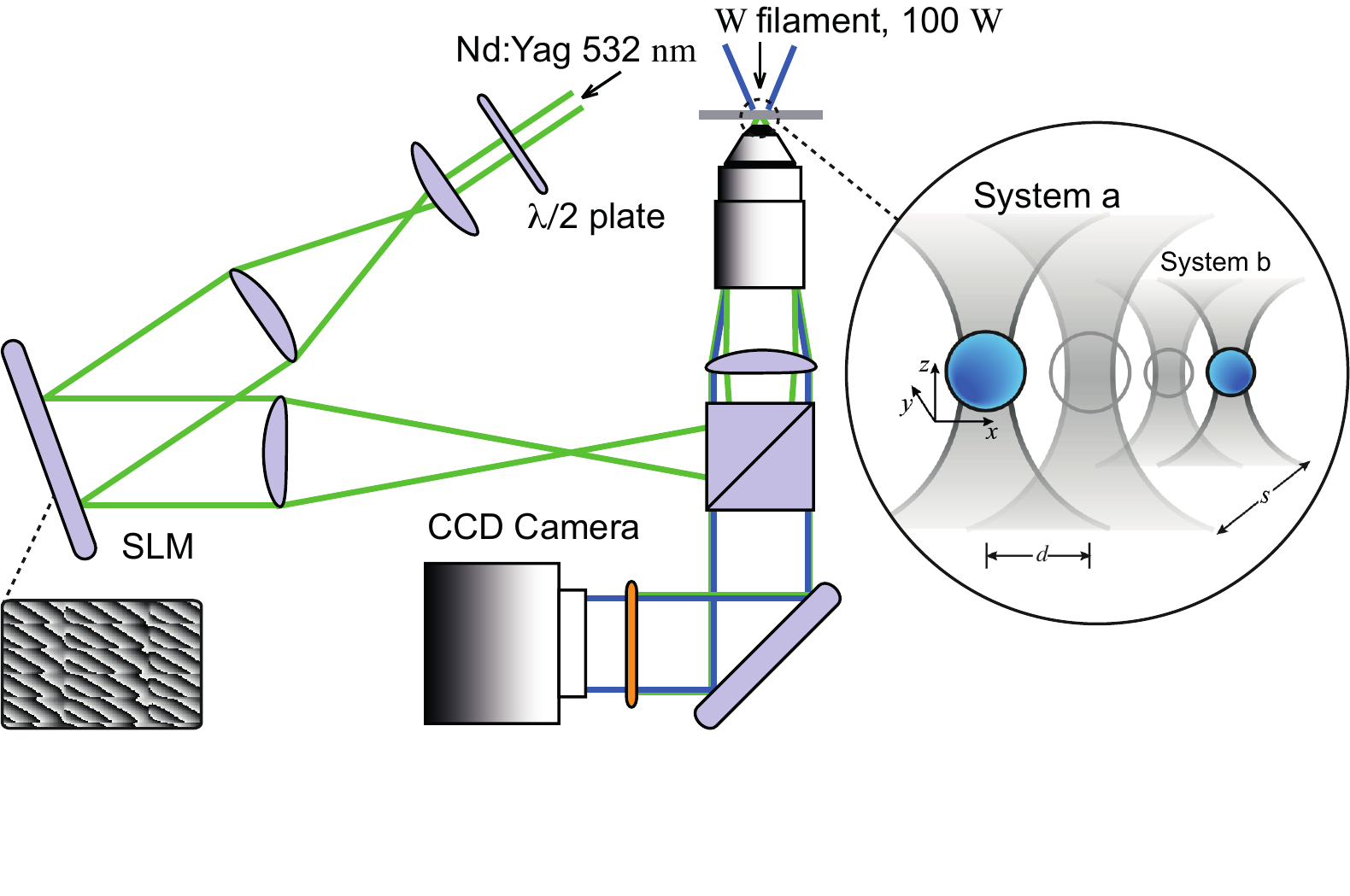}
\caption{Holographic optical tweezers. A spatial light modulator is used to generate multiple traps at the focal plane of the $\times100$ oil-immersion microscope objective. Silica spheres are trapped with a frequency-doubled Nd:Yag laser, $5$ $\mu m$ above the cover slip. Left inset shows the hologram used to landscape two side by side bistable potentials. A schematic of the two optically trapped spheres in independent bistable potentials is inset on the right.}
\label{FIG1}
\end{figure}

Sculpting of the bistable optical landscape is achieved by a holographic phase element generated by a gratings and lenses algorithm \cite{Liesener_algo_2000,Leach_algo_2006}. Holograms are imaged at the back aperture of a Carl Zeiss $\times100$ $1.3$ NA oil-immersion microscope objective and focused into a multispot array located at the focal plane (Fig. \ref{FIG1}). Each bistable potential is constructed by having two individual focal spots placed in close proximity, such that a surmountable barrier forms between the two traps. We create two bistable potentials separated by a distance $s$, which is also controlled holographically.  Care is taken to maintain symmetric bistable potentials by adjusting the optical intensities of each trap. Stilgoe \emph{et al.} have recently given a  report on the complexity of creating optical potentials using two optical traps \cite{stilgoe:2011p248101}. Our samples are prepared with Silica spheres (of radius, $a=0.4 \mu$m, Bangs Laboratories) in de-ionized, distilled water with a 1:10$^{6}$ ratio. When preparing our samples we add a weak saline solution just before sealing so that any free debris is immobilised on the cover slip due to the reduced Debye length \cite{Alvarez:2010p5911} whilst still allowing enough time to capture the two spheres required for the experiment. Sample cells are constructed with a single concave microscope slide (concave depth $= 500$ $\mu m$) and a square cover slip (thickness number $1.5$), sealed with UV-curing optical adhesive.  Using a few $mW$ of frequency-doubled Nd:Yag laser light ($532$ $nm$), the trap stiffness in each trap is set at $\sim 1 pN/ \mu m^{-1}$. A Prosilica GE650 camera, operating at a frame rate of 1.55 $kHz$ is used to track the position of the spheres in the $x-y$ plane. Using a center of mass algorithm, these in-plane displacements are measured  with $nm$ precision \cite{Padgett_Gibson}. Time series of the spheres' center of mass, $x(t)$ and $y(t)$, are extracted in blocks of $10^5$ frames and accumulated for a total time period of $2$-$3$ hours, significantly longer than the average between hops, which is of order one second. 

\begin{figure}[htb]
\includegraphics[width=8cm]{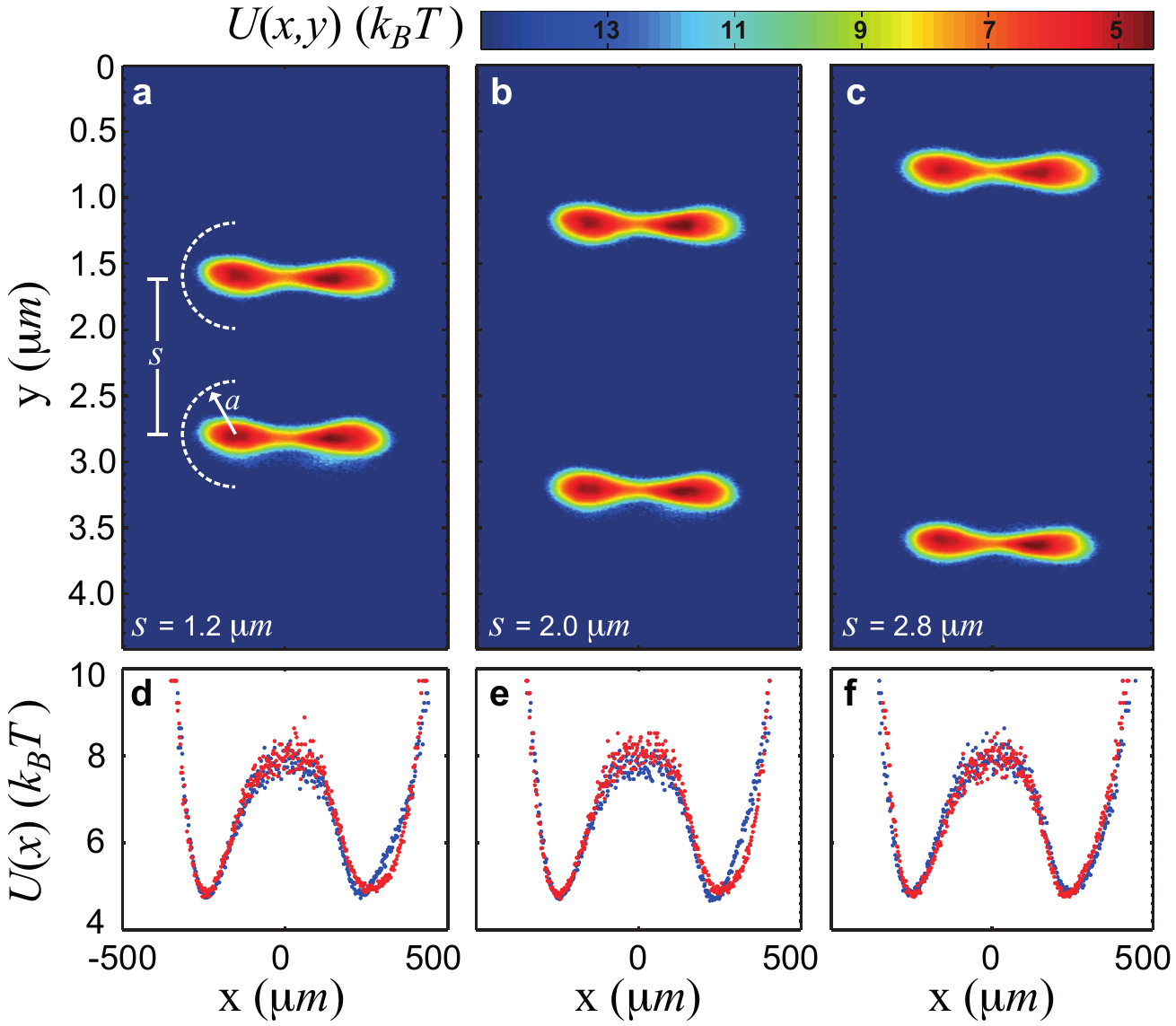}
\caption{Optical potential landscapes experienced by two $800$ $nm$ Silica spheres. Potential energy distributions are recovered from particle trajectories, $x(t)$ and $y(t)$, as described in the text. The data is binned into bins of size $10$ $nm^2$ for a-b and $2$ $nm$ for d-f.}
\label{FIG2}
\end{figure}

\begin{figure}[htb]
\includegraphics[width=8cm]{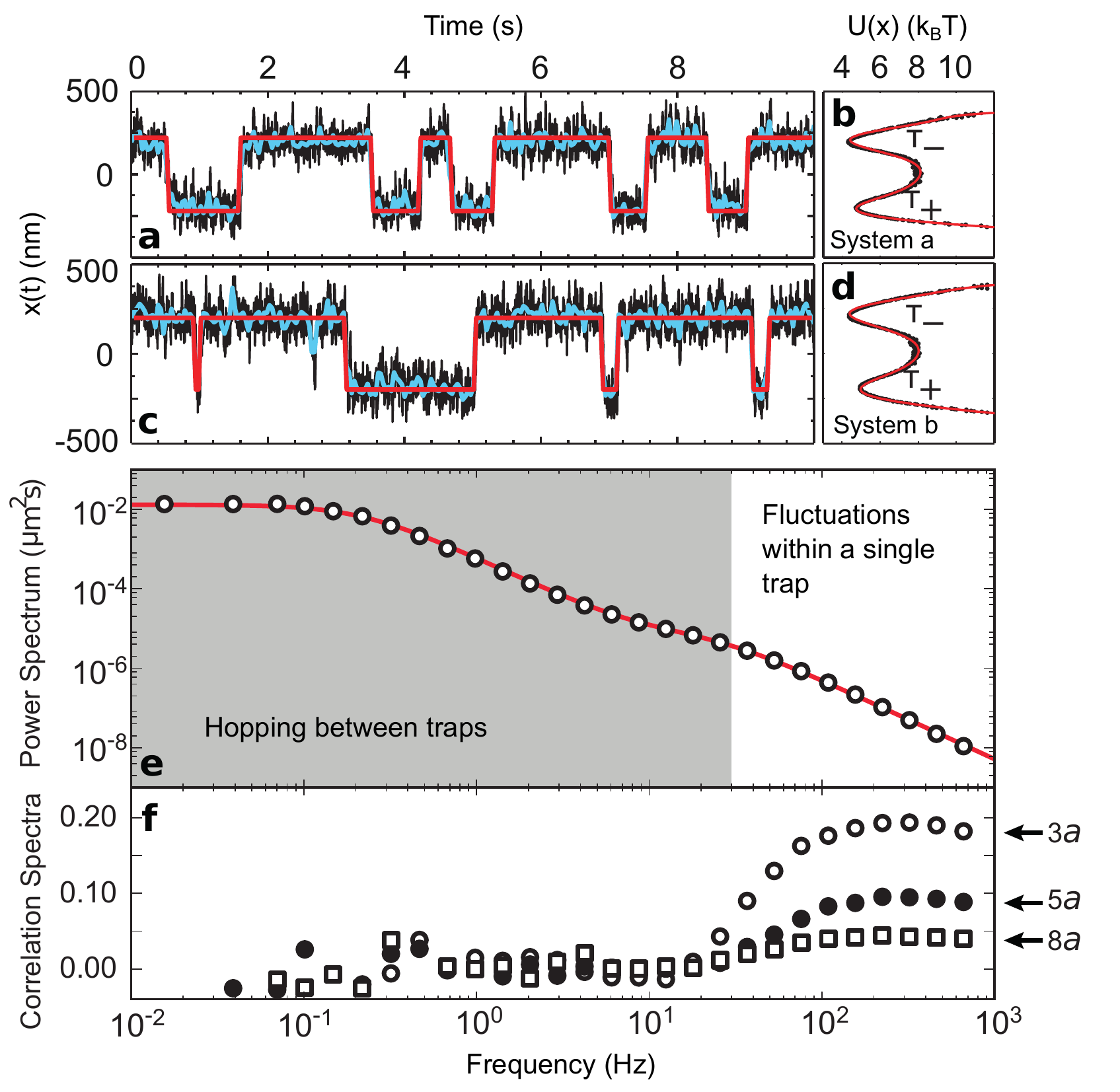}
\caption{(a, c) $10$ $s$ sample of the spheres' position along the $x$-axis (black line) for each system separated by $s$ $=$ $3.6$ $\mu m$. The block filtered (blue line) and digitised traces (red line) are also shown. (b, d) Corresponding potentials (black $\circ$) with fitted polynomials (red line). (e) Single particle power spectrum for $s$ $=$ $1.2$ $\mu m$ along with a double Lorentzian fit as discussed in the text. (f) Correlation spectra for three separations evidencing hydrodynamic correlations in the high frequency region.}
\label{FIG3}
\end{figure}

Since the sampling rate is much greater than the corner frequency in the power spectrum of particle fluctuations, $f_c\sim20$ $Hz$, the normalized histogram of particle positions $\rho(x,y)$ provides a direct measurement of the Boltzmann distribution, $\rho(x,y) = Z^{-1} \exp[-U(x,y)/k_B T]$, where $k_B$, $T$ and $Z$ are respectively the Boltzmann constant, the temperature of the surrounding fluid and the normalisation constant. By inverting $\rho(x,y)$ we can directly access the underlying optical potential in units of $k_B T$ (Fig. \ref{FIG2} and \ref{FIG3} (b, d)).

The obtained potentials are shown in Figure \ref{FIG2} (a - c)  for three  separations, $s$ $=$ $1.2$, $2.0$ and $2.8$ $\mu m$. The potentials along the hopping axis for system $a$ (blue $\circ$) and system $b$ (red $\circ$), presented in Fig. \ref{FIG2} (d-f), confirm that the optical landscapes of each system are not significantly perturbed as they approach. It is worth noting that, although the optical landscape is made of  two optical traps whose centers are separated by $800$ $nm$, the separation between the minima of the obtained bistable potential is only $d \sim$ $400$ $nm$. This is due to the sphere experiencing restoring forces from both optical traps, leading to an offset in the equilibrium position in each individual trap and shortening the hopping distance.

Figure \ref{FIG3} (a, c) shows a $10$ $s$ sample trace of particle position on the $x$-axis (the axis along which hopping occurs) taken at large separation, $s$ $=$ $3.6$ $\mu m$. The corresponding potentials are given in Fig. \ref{FIG3} (b, d). A single particle power spectrum is shown in Fig. \ref{FIG3} (e) for the shortest separation, $s$ $=$ $1.2$ $\mu m$. Two separated timescales are visible, a low frequency component describing the hopping dynamics and a high frequency component describing the fluctuations around the local minima. For an isolated hopping system, the Fourier amplitude, $\hat x(\nu)$, of the particle's $x$ position will be distributed with a double Lorentzian spectrum,

\begin{equation}
\langle \hat x^*(\nu)\hat x(\nu)\rangle =  
\frac{k_B T}{\pi k}\frac{f_c}{\nu^2+f_c^2}+
\frac{d^2}{4\pi}\frac{2 f_h}{\nu^2+(2 f_h)^2}.
\end{equation}

\noindent
The first term is the standard spectrum for a bead of mobility $m_0=1/(6 \pi \eta a)$ (where $\eta=1mPas$ is taken as the viscosity) in an optical trap of strength $k$, and thus, a corner frequency given by $f_c=m_0 k/2\pi$. The second term describes hopping events over a distance $d$ and a jump probability per unit time given by $f_h$. The hopping frequency can be anticipated using Kramers theory $f_h=\omega_0 \exp[-\Delta U/k_B T]$ where $\Delta U$ is the height of the potential barrier and the prefactor $\omega_0=m\sqrt{\kappa_i \kappa_s}$ depends on the particle mobility $m_0$ and on the curvatures of the underlying potential at the stationary points $\kappa_{i,s}=\left|\partial^2 U/\partial x^2\right|_{x=x_{i,s}}$. 
Knowing the temperature, viscosity and particle radius means that all remaining parameters can be directly evaluated from the features of the stationary points in the obtained potential energy. The resulting curve is shown as a solid line in Fig. \ref{FIG3} (e) and fits the measured power spectrum well. Searching for hydrodynamic correlations we plot in Fig. \ref{FIG3} (f) correlation spectra $\langle \hat x_a^*(\nu) \hat x_b(\nu)\rangle$ normalized by the average single particle power spectra $(1/2)\sum_{\alpha=a,b}\langle \hat x_\alpha^*(\nu) \hat x_\alpha(\nu)\rangle$. A marked positive correlation at high frequencies gets larger as the two systems approach. This is the well known phenomenon by which the mobility of rigid motions is higher than that for relative motions so that at short times the particles tend to move collectively \cite{Meiners:1999p2211}. However, at low frequencies dominated by hopping dynamics, no sign of correlation is observable even when the two systems are close together. 

Despite the lack of any obvious sign of collective motion in the low frequency correlations spectrum (Fig. \ref{FIG3} (f)), further analysis of our data reveals a significant correlation. Rather than averaging the correlation over all points, we examine only the instances where both particles hop at the same time, either in the same direction (symmetric) or opposite direction (antisymmetric).  To examine these instances we first average the position time series in blocks of 60 frames (Fig. \ref{FIG3} (a, c) blue line). This acts as a low pass filter with a cutoff frequency of about $25$ $Hz$, just higher than the trap corner frequency, filtering out the fast dynamics of fluctuations during a hop.
The blocked data is then digitized into a two state variable $ \bar{x}( t ) \in { -1 , +1 } $, denoting the left and right states of the bistable potential (Fig. \ref{FIG3} (a, c) red line). For each system we then extract a time series of hop events from $-1 \rightarrow +1$  and $+1 \rightarrow -1$.   This data represents a cumulative acquisition of $215$ minutes at $1550$ $Hz$, with $\sim$ $10^4$ transition events. The distributions of intervals between hops are fitted with an exponential decay, yielding the characteristic dwell time for each state. These measured dwell times are consistent with Kramers formula applied to the detailed shape of the underlying potential. 
Each bistable potential is fitted with an eighth order polynomial (Fig. \ref{FIG3} (b, d) red line), the second derivative of which gives the Kramers prefactor term $\omega_0$ when evaluated at the stationary points. Numerical results of $\omega_k^{-1}$ are given in the third column of Table \ref{tab}. The dwell times are measured from $x(t)$, the mean of which are given in the second column. The probability distribution of the dwell times are fitted with an exponential decay giving the first column in Table \ref{tab}. 

\begin{table}
\begin{center}
\renewcommand{\arraystretch}{1.2}
\begin{tabular}{ ll | llll ccc cc }
		&&&&	Fit		&&&	Mean	&&&	$\omega_k^{-1}$	\\ \hline
System a	&&&&	0.54		&&&	0.57		&&&	0.46			\\ 
		&&&&	0.66		&&&	0.68		&&&	0.62			\\
System b	&&&&	0.48		&&&	0.50		&&&	0.37			\\
		&&&&	0.74		&&&	0.73		&&&	0.61			\\
\end{tabular}
\end{center}
\caption{Characteristic dwell times for each individual potential obtained by exponential fitting to the probability distributions of the states dwell times, along with the mean value of raw dwell times and from Kramers formula. All numbers are in units of $s$.}
\label{tab}
\end{table}

\begin{figure}[htb]
\includegraphics[width=8cm]{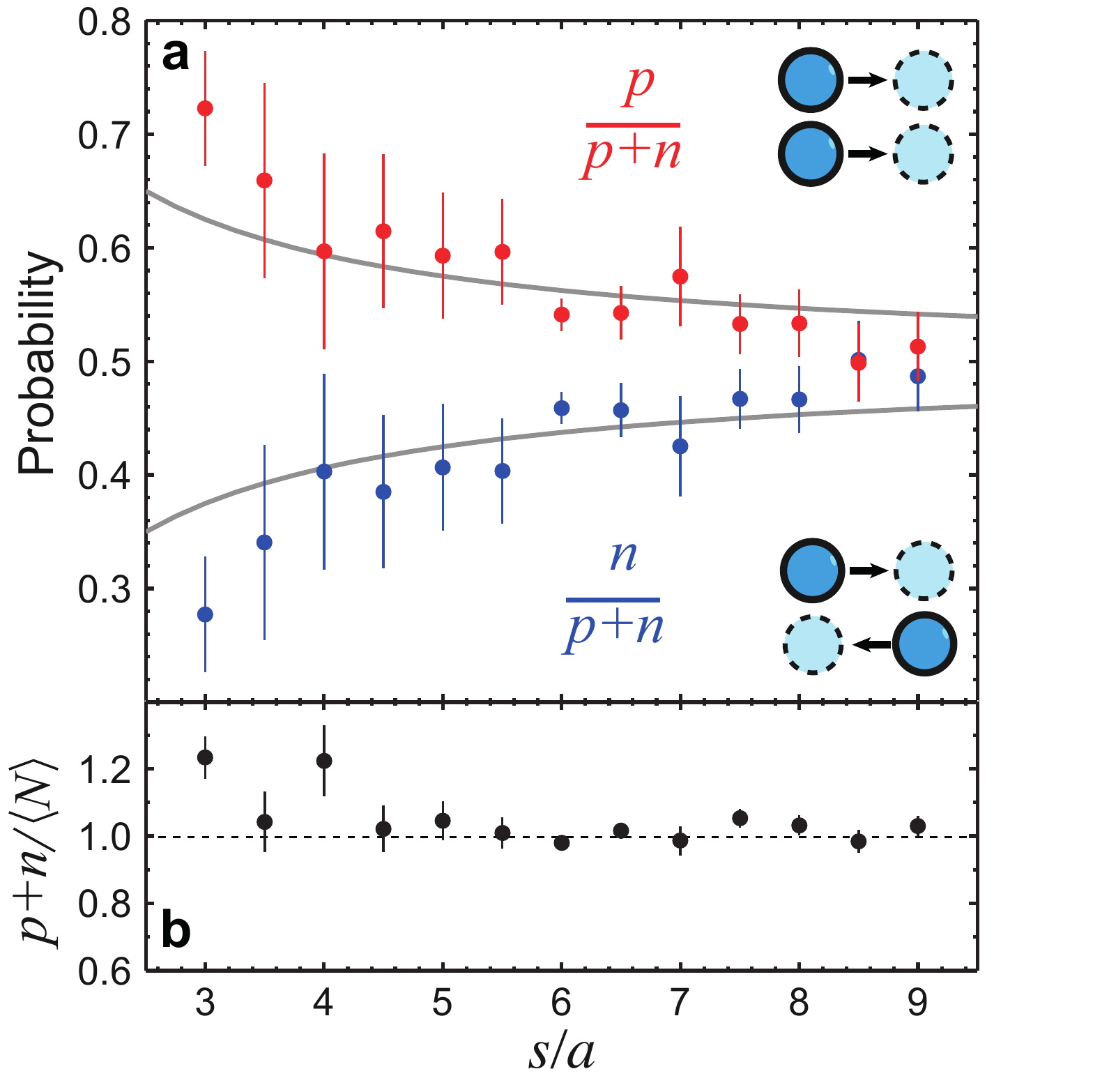}
\caption{(a) Probability of coincidences over a range of system separations, $s$. $p$ and $n$ coincidences are defined by the inset cartoon. Coincidence events are counted as described in the text and presented as red $\circ$ for $p$ and blue $\circ$ for $n$. Data is normalised by $p+n$. The solid lines represent the splitting in $p$ and $n$ assuming a linear dependence on hydrodynamic coupling with a strength of $3a/4s$. (b) The total counted coincidences at each system separation normalised by the expected number, $\left<N\right>$, (Eq. \ref{eq:expectedN}) with an error of $\pm \sqrt{p+n}/\left<N\right>$.}
\label{FIG4}
\end{figure}

Each transition event is assigned a duration of one blocked sequence of frames, $\delta t =$ $39$ $ms$. We count the number of occurrences of possible coincident events, namely $p$ and $n$, where we define a $p$ coincident as having occurred when both spheres hop in the same direction and an $n$ coincidence when both spheres hop in opposite directions. Figure \ref{FIG4} (a) shows this nomenclature. It is expected that two uncoupled stochastic oscillators will show a finite number of coincidences, $\langle N \rangle$, depending on the observed time period, $T$, and the temporal window, $\delta t$, in which we define a coincidence and the transition rates of the two oscillators, $\tau_{a,b}^{-1}$, given by,

\begin{equation}
\langle N \rangle = \delta t T \left(\tau_a \tau_b \right) ^{-1}.
\label{eq:expectedN}
\end{equation}

For two uncoupled stochastic oscillators we expect an equal probability for symmetric and antisymmetric simultaneous hops ($p=n$). Figure \ref{FIG4} (a) shows the measured probabilities of $p$ and $n$ coincidences occurring over a range of system separations. With the two systems placed in close proximity to each other ($s \sim 3a$) the probability of observing a $p$ coincidence is significantly greater than observing a $n$ coincidence ($p - n > \langle N \rangle^{-\frac{1}{2}}$). At increasing separations, both $p$ and $n$ coincidence probabilities approach $0.5$ ruling out the possibility that the synchronisation could be ascribed to asymmetries in the optical landscapes.   We note that the total number of observed coincidences, $(p+n) /\langle N\rangle$, is unchanged with separation (Fig. \ref{FIG4} (b)).  Coming back to Kramers' formula for single particle hopping rate, we notice that particle mobility appears as a prefactor to an energetic activation term. We may then conjecture that the rates of symmetric hops and antisymmetric hops have a similar form with an activation term and a prefactor that would also depend respectively on the collective and relative mobilities \cite{Dufresne:2000p187,curran_2010,LowRey_Book}.  Moreover, the splitting between $p$ and $n$ seems to decay with distance as is the case for hydrodynamic splitting between collective and relative mobilities as shown by the $0.5\pm(3/4)(a/s)$ lines in Fig. \ref{FIG4} (a).


In conclusion, the thermally activated jumps of two $800$ $nm$ silica spheres in neighboring bistable optical landscapes are shown to be coupled via hydrodynamic interactions.  Due to the higher mobility of collective motions, when two systems are in close proximity, there is a higher probability of observing a symmetric hop whilst antisymmetric hops are less common. We argue that the experimental environment studied here provides an idealised representation of interacting stochastic oscillations that occur in nature. It will be interesting to extend the present study to a larger ensemble of bistable systems. For example, looking for the emergence of more complex, cooperatively rearranging regions could aid the understanding of the role of hydrodynamic interactions in the glassy dynamics of concentrated colloidal suspensions \cite{Weeks:2000p627}.

\end{document}